\documentclass[aps,twocolumn,
superscriptaddress,
footinbib,
prx]{revtex4-2}

\usepackage{amsmath,amssymb,bm}
\usepackage{graphicx}
\usepackage{epstopdf}
\usepackage{latexsym}
\usepackage[normalem]{ulem} 
\usepackage[caption=false, position=top]{subfig}
\usepackage[usenames,dvipsnames]{color}
\usepackage{hyperref}
\usepackage{natbib}
\usepackage{xcolor}
\usepackage{physics}
\usepackage{multirow}
\usepackage{stackengine}
\usepackage{dsfont}
\usepackage{bbold}
\usepackage{nicefrac}
\usepackage[inkscapelatex=false]{svg}
\usepackage{enumitem}
\usepackage{tikz}
\usetikzlibrary{quantikz}

\usepackage{framed}

\usepackage[utf8]{inputenc} 

\newcommand{\comment}[1]{}
\newcommand{\taulink}[1]{\tau_{#1, #1 + 1}}

\newcommand{\g}{g}

\def\equationautorefname~#1\null{Eq.~(#1)\null}
\begin{document}

\title{Stabilizing quantum simulations of lattice gauge theories by dissipation}

\date{\today}

\author{Tobias Schmale}
\email{tobias.schmale@itp.uni-hannover.de}
\affiliation{Institut f\"ur Theoretische Physik, Leibniz Universit\"at Hannover, Appelstraße 2, 30167 Hannover, Germany}
\author{Hendrik Weimer}
\email{weimer@tu-berlin.de}
\affiliation{Institut f\"ur Theoretische Physik, Technische Universit\"at Berlin, Hardenbergstraße 36 EW 7-1, 10623 Berlin, Germany}
\affiliation{Institut f\"ur Theoretische Physik, Leibniz Universit\"at Hannover, Appelstraße 2, 30167 Hannover, Germany}

\begin{abstract}
  Simulations of lattice gauge theories on noisy quantum hardware
  inherently suffer from violations of the gauge symmetry due to
  coherent and incoherent errors of the underlying physical system
  that implements the simulation. These gauge violations cause the
  simulations to become unphysical requiring the result of the simulation
  to be discarded. We investigate an active correction scheme
  that relies on detecting gauge violations locally and subsequently
  correcting them by dissipatively driving the system back into the
  physical gauge sector. We show that the correction
  scheme not only ensures the protection of the gauge symmetry, but it also
  leads to a longer validity of the simulation results even within the
  gauge-invariant sector. Finally, we discuss further applications of
  the scheme such as preparation of the many-body ground state of the
  simulated system.
  \end{abstract}

\maketitle The quantum simulation of lattice gauge theories is
one of the most important applications of quantum computers, given
that the fundamental field theories that govern modern physics are
gauge theories. However, error processes in noisy intermediate-scale
(NISQ) quantum devices may violate the gauge symmetry that is
essential for the viability of the simulation. Here, we show that
engineered dissipation can overcome these limitations and protect the
validity of the results for much longer simulation times.

The main challenge when implementing a quantum simulation of a lattice
gauge theory is that the gauge symmetry has to be explicitly
programmed into the device \cite{Tewari2006,Weimer2010,Wiese2013,Stannigel2014,Dalmonte2016}. For current NISQ
devices, this means that unavoidable errors will generically lead to
violations of the gauge symmetry and result in the creation of quantum
states that are unphysical simulation results. To solve this problem,
several solutions have been offered. One solution is to integrate out
the redundant degrees of freedom. This however typically results in
non-local interactions \cite{martinez_real-time_2016}, increasing the
complexity of the simulation. Other approaches include modifications
to the Hamiltonian to attribute an energy penalty to a gauge
violation \cite{halimeh_gauge_symmetry_2021}, adding (random) gauge
transformations during time evolution in order to stochastically
cancel gauge errors \cite{tran_faster_2021, lamm_suppressing_2020} or
detecting gauge violations using oracles and rejecting simulations
with gauge violations \cite{stryker_oracles_2019}.

In our work, we build on the vast work on the realization of
engineered dissipation channels for quantum many-body systems
\cite{Diehl2008,Verstraete2009,Barreiro2011,Krauter2011,Carr2013a,Rao2013,Lin2013,Shankar2013,Morigi2015,Reiter2016,Weimer2017,Roghani2018,Metcalf2020,Jamadagni2021,Piroli2021}. Specifically, we detect gauge violations during the time
evolution and apply local correction operations in real time to fix
any gauge violations that might occur. We investigate the effects of
our scheme on the accuracy of simulations and show control over the
temperature of the system as an additional benefit of our approach.

\subsection{Model gauge theory}
For concreteness, we focus on a paradigmatic $\mathbb{Z}_2$ lattice gauge theory as one of the simplest, yet non-trivial lattice gauge theories \cite{halimehFateLatticeGauge2020}. This is a relevant test-bed for our setup, as $\mathbb{Z}_2$ gauge theories naturally appear in digital quantum simulations of fermionic lattice models \cite{verstraete_mapping_2005, kitaev_anyons_2006} as well as in the realization of paradigmatic models of topological order \cite{kitaev_fault-tolerant_2003}. We will address possible generalizations of our setup to other gauge theories in the next section.

The $\mathbb{Z}_2$ Hamiltonian we study is given by
\begin{align}
    H_0 = \sum_{j=1}^N \left[J_a(\sigma^+_j\taulink{j}^z\sigma^-_{j+1} + \text{h.c.}) - J_f \taulink{j}^x\right], \label{eq:hamiltonian_base}
\end{align}
which acts on $N$ matter sites, each of which can contain either the vacuum or a hard-core boson. These are described by the Pauli ladder-operators $\sigma^{+/-}$. In between the matter sites sit gauge link variables $\taulink{j}^{x/z} = \sigma^{x/z}_{j, j+1}$, represented by Pauli $x/z$ matrices, where the label $j, j+1$ represents the matter sites adjacent to the gauge site. Therefore, the system can be mapped to $2N$ qubits. Following \cite{halimehFateLatticeGauge2020}, we use periodic boundary conditions and set the matter-field coupling $J_a = 1$ and the electric field energy $J_f = 0.54$.
Gauge-invariance of this Hamiltonian is defined by the Gauge operators
\begin{align}
    G_j =1 - (-1)^j\tau_{j-1, j}^x\sigma^z_j\taulink{j}^x, \label{eq:gauge_operators}
\end{align}
which satisfy $[H_0, G_j] = [G_j, G_l] = 0\ \forall j, l$. Once initialized in an eigenvalue $g_0$ of the gauge operators, the time-evolution should therefore not change this eigenvalue. 

However, during a simulation of such a lattice gauge theory on a quantum simulator, unitary and non-unitary errors can break gauge invariance, leading to gauge eigenvalues $g$ deviating from the initial $g_0$. We incorporate these unitary effects by adding a small gauge variant perturbation 
\begin{align}
    H_1 &= \sum_{j=1}^N \left[(\sigma_j^+ (c_1 \taulink{j}^- +c_2 \taulink{j}^+)\sigma^-_{j+1} + h.c.)\right]\\& + \sum_{j=1}^N \sigma^+_j\sigma^-_j(c_3\taulink{j}^z + c_4\tau^z_{j-1, j}) \label{eq:hamiltonian_gauge_violating}
\end{align}
to the Hamiltonian, resulting in the complete Hamiltonian
\begin{align}
    H = H_0 + \lambda H_1. \label{eq:hamiltonian_full}
\end{align}
The precise values of the dimensionless coupling constants $c_i$ are not important, in line with previous work we set them to $c_1 = 0.51$, $c_2 = -0.49$, $c_3 = 0.77$, $c_4=0.21$ \cite{halimehFateLatticeGauge2020}.
Incoherent errors are represented by single-site Lindbladian jump operators $L_i$ acting on all matter and link sites, resulting in the model dynamics being described by the Lindblad master equation
\begin{align}
  \dot{\rho} = i[\rho, H] + \sum_i \left[L_i\rho L_i^\dagger - \frac{1}{2} \rho L_i^\dagger L_i - \frac{1}{2} L_i^\dagger L_i\rho \right].
  \label{eq:lindblad}
\end{align}
As $L_i$ we investigate bit-flips $\sqrt{\gamma}\sigma^x$ and phase-flips $\sqrt{\gamma}\sigma^z$ on matter and link sites, and alternatively spontaneous emission $\sqrt{\gamma}\sigma^- = \sqrt{\gamma}(\sigma^x - i \sigma^z\sigma^x)$ on all sites.

We quantify the gauge violation by 
\begin{align}
    \varepsilon (t) &= \frac{1}{N}\left|\sum_{j=1}^N\expval{G_j(t)} - \sum_{j=1}^N\expval{G_j(0)}\right| =  \frac{1}{N}\sum_{j=1}^N\expval{G_j(t)},\label{eq:gauge_violation}
\end{align}
where the last equation holds by the choice of the initial gauge $\expval{G_j(0)} = 0\ \forall j$.

\subsection{Active gauge correction scheme}

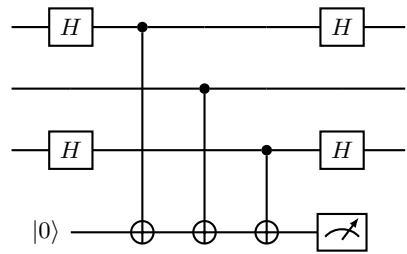
\begin{figure}[h!]
\centering
\begin{quantikz}
 &\gate{H}          &\ctrl{3}&\qw     &\qw     &\gate{H}&\qw\\
 &\qw               &\qw     &\ctrl{2}&\qw     &\qw     &\qw\\
 &\gate{H}          &\qw     &\qw     &\ctrl{1}&\gate{H}&\qw\\
 &\lstick{$\ket{0}$}&\targ{} &\targ{} &\targ{} &\meter{}&
\end{quantikz}
\caption{Circuit for performing the ancilla assisted stabilizer measurement necessary to detect a gauge violation. The three qubits correspond to the three sites that the gauge operator acts on.} \label{fig:measurement_implementation}
\end{figure}

We imagine the above lattice gauge theory to be simulated in a Trotterized manner on a digital quantum computer, where time-evolution is split into small time-steps, which are implemented by unitary operations \cite{Lloyd1996,martinez_real-time_2016, smith_simulating_2019, fauseweh_quantum_2024}. After a variable number of time-steps, mid-circuit measurements of all gauge operators can be performed, e.g. using the circuit from Fig.~\ref{fig:measurement_implementation}, making use of the fact that the gauge operators and the Hamiltonian all mutually commute. This yields a sequence of gauge eigenvalues corresponding the each gauge operator $(G_1, ..., G_N)$. We can now react to this measured gauge syndrome and apply unitary operations to restore the original gauge $(G_1, ..., G_N) = (0, ..., 0)$. 

In order find a set of possible correction operators, we first study the action of bit-flips and phase flips on the gauge eigenvalue. The gauge operators $G_j$ have two eigenvalues $0$ and $2$. A bit-flip on matter site $j$ will flip the eigenvalue of $G_j$. Similarly, a phase-flip on link site $j$ will flip the eigenvalue of the operators $G_j$ and $G_{j+1}$. On the other hand, bit-flips on link sites and phase-flips on matter sites do not result in a change of the gauge eigenvalue. This is summarized in Table~\ref{tab:gauge-syndromes}. It therefore makes sense to separate the errors into \textit{gauge-variant}, i.e. gauge violating errors and the remaining \textit{gauge-invariant} errors. 

\begin{table}[h]
    \centering
    \begin{tabular}{c|c}
    Error & Gauge eigenvalues \\ & $G_{j-i}, G_j, G_{j+1}$\\\hline
      $\sigma^x_j$   &  0, 2, 0\\
      $\sigma^z_j$   &  0, 0, 0\\
      $\taulink{j}^x$   &  0, 0, 0\\
      $\taulink{j}^z$   &  0, 2, 2
    \end{tabular}
    \caption{Summary of how single qubit errors affect the gauge.}
    \label{tab:gauge-syndromes}
\end{table}

Since not all errors are visible to the gauge operators, one can not hope to achieve full error correction from the gauge theory alone. It is possible to extend the lattice gauge theory to a full quantum error correcting code by adding more degrees of freedom and formulating constraints on these \cite{rajput_quantum_2023}. Instead we here investigate phenomena that to not require extra qubit resources.

The gauge syndromes in Table~\ref{tab:gauge-syndromes} suggests the following correction scheme:
\begin{itemize}
    \item If a sequence $(G_{j-i}, G_j, G_{j+1}) = (0, 2, 0)$ is measured, a correcting bit-flip is applied to matter site $j$. 
    \item If a sequence $(G_j, G_{j+1}) = (2, 2)$ is measured, a correcting phase-flip is applied to link site $j$. 
\end{itemize}
In both cases, all gauge eigenvalues are subsequently restored to the physical sector after correction.
This scheme could now be applied stroboscopically after a fixed number $n$ of Trotter steps, each evolving the system by time $dt$, using mid-circuit measurements and real-time feedback \cite{koh_measurement-induced_2023, schmale_real-time_2023}.\\
For our numerical simulations, we instead choose a different but equivalent formulation based on jump operators.
We define $P^0_j$ and $P^2_j$ as the projectors that project into the two eigenvalues $0$ and $2$ of $G_j$. In fact, here this means that $P^2_j = G_j/2$ and $P^0_j = 1 - P^2_j$.
We then define the \emph{correction jump operators} as 
\begin{align}
    C^x_j &= \sqrt{\gamma_c} \sigma^x_j P^0_{j-1}P^2_jP^0_{j+1} \qquad \text{and}\\
    C^z_j &= \sqrt{\gamma_c} \taulink{j}^z P^2_jP^2_{j+1},
\end{align}
which correct the two gauge syndromes given above and annihilate any
other sequence of gauge eigenvalues. $\gamma_c = \frac{1}{n\cdot dt}$
is the \emph{correction rate} and specifies how many correction
operations are applied per unit time.  We then simulate the dynamics
of the correction scheme by using these operators as additional jump
operators in the master equation. This is equivalent to the
stroboscopic application of the correction circuit since the latter
corresponds to a Trotterized form of the master equation dynamics
\cite{Weimer2010}.

\begin{figure}[h!]
    \centering
    \includegraphics[width=0.5\textwidth]{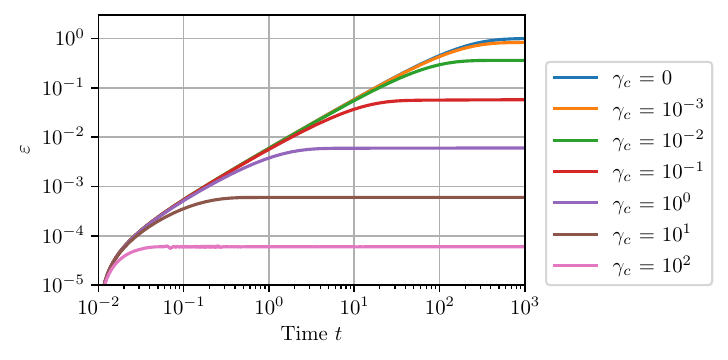}
    \caption{The gauge violation \eqref{eq:gauge_violation} is tracked over time for different correction rates $\gamma_c$, starting from the ground-state ($\lambda=10^{-2}$, $\gamma=10^{-3}, N=4$). Higher correction rates result in smaller gauge violations.}
    \label{fig:gauge-correction}
\end{figure}

The result of this correction scheme in action can be seen in Figure~\ref{fig:gauge-correction}. Without gauge correction ($\gamma_c = 0$), the gauge violation $\varepsilon$ increases linearly until a steady-state of maximal gauge violation is reached. Turning on the correction ($\gamma_c > 0$) results in this steady-state violation to be suppressed inversely proportional to the correction rate $\gamma_c$. This shows that this scheme is effective in suppressing gauge errors and therefore prevents the simulation of the lattice gauge theory to turn unphysical.  

This measurement scheme can be extended to more general gauge theories such as $\mathbb{Z}_n$ and $U(1)$ as shown in \cite{stryker_oracles_2019}, 
while a possible generalization of the corrections can be found in \cite{rajput_quantum_2023}.
We note that since the gauge operators mutually commute, the physical gauge sector can be written as a ground space of a local stabilizer Hamiltonian and hence our scheme implements a form of stabilizer pumping  \cite{Weimer2010,Weimer2011,mullerSimulatingOpenQuantum2011}. For non-Abelian gauge theories, the physical gauge sector is instead the ground space of a local, frustration-free Hamiltonian \cite{Banerjee2013}, for which efficient preparation strategies are known as well \cite{Verstraete2009}. However, the study of these methods in the context of gauge-correction is outside the scope of this work.

\subsection{Sympathetic cooling during gauge correction} 

We now make a slight modification to the Hamiltonian by adding the gauge violation as a new term
\begin{align}
    H \rightarrow H + \g \frac{1}{N}\sum_{j=1}^N G_j. \label{eq:gauge_penalty}
\end{align}
This associates an energy penalty $\g$ to a gauge violation. We choose
$g=1$ and show a dependence of our results on this choice in the
appendix. While such a term can be used to reduce gauge violations
stemming from \emph{coherent} errors such as in $H_1$
\cite{halimeh_gauge_symmetry_2021}, they cannot directly suppress
\emph{incoherent} errors such as the ones in the master equation
(\ref{eq:lindblad}). However, the combination with our dissipative
gauge correction scheme realizes a setup corresponding to a sympathetic cooling of the gauge-invariant sector \cite{raghunandanInitializationQuantumSimulators2020}. Here, the gauge-invariant sector is the system to be cooled, while the gauge degrees of freedom act as a bath. The gauge degrees of freedom are rapidly cooled into the ground state of having no gauge violations, while the coherent gauge errors in Eq.~\eqref{eq:hamiltonian_gauge_violating} lead to a system-bath coupling and allow energy to be dissipated out of the gauge-invariant sector.

This is similar to the cooling schemes shown in
\cite{raghunandanInitializationQuantumSimulators2020} and
\cite{polla_quantum_2021, mi_stable_2024}, with the crucial difference
that these schemes require extra bath degrees of freedom to be
artificially added to achieve cooling, therefore increasing the cost
of the simulation. Here, the bath is implemented by the gauge degrees
of freedom and does not require any additional resources.

\begin{figure}[h!]
    \centering
    \includegraphics[width=0.5\textwidth]{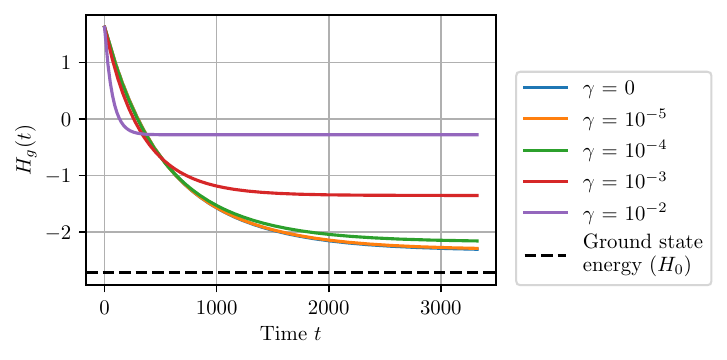}
    \caption{Cooling is demonstrated by the gauge sector energy \eqref{eq:gauge_sector_energy} decreasing during time evolution. The steady state energy depends on the incoherent error rate $\gamma$ ($\lambda=0.1, \gamma_c=1, N=3$, $\sigma^-$ decay). }
    \label{fig:cooling}
\end{figure}

\begin{figure}[h!]
    \centering
    \includegraphics[width=0.5\textwidth]{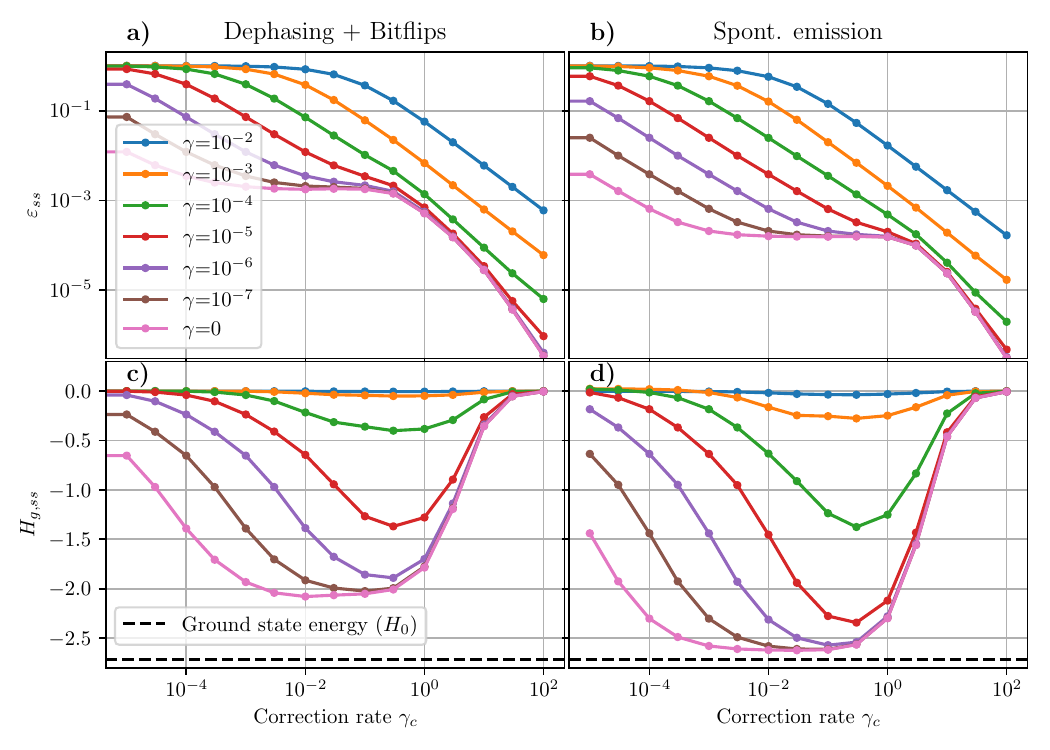}
    \caption{The steady state gauge violation (\textbf{a)} and \textbf{b)}) from Eq.~\eqref{eq:gauge_violation} and gauge sector energy (\textbf{c)} and \textbf{d)}) from Eq.~\eqref{eq:gauge_sector_energy} is shown for various correction rates $\gamma_c$ and errors with rate $\gamma$ on all sites. Maximum cooling is achieved in intermediate correction regimes where correction rates are not too small to be effective and not too large to fall into the Zeno regime, as explained in the main text (\textbf{a)} and \textbf{c)}: Dephasing and bit-flips on all sites, \textbf{b)} and \textbf{d)}: Spontaneous emission on all sites, $\lambda=0.03, N=3$).}
    \label{fig:steady_xz_errors_vs_gammac}
\end{figure}

We show this in Figure~\ref{fig:cooling}, where we start in a high-energy product state and demonstrate how the dissipative time-evolution drives the system to lower energy states. 
To ensure that any energy difference is not a direct consequence of the gauge penalty \eqref{eq:gauge_penalty}, we only measure the energy of the system in the physical sector as
\begin{align}
    H_g(t) = \frac{\text{Tr}[PH_0P \rho(t)]}{\text{Tr}[P\rho(t)]}, \label{eq:gauge_sector_energy}
\end{align}
where $P=\prod_j P^0_j$ is the projector into the physical gauge sector.
Depending on the magnitude of the competing heating caused by the dissipative errors, an energy close to the ground state can be reached. 
This final state is then stable under further time evolution, i.e. is protected from further heating. We show this by directly solving for the steady state of the time evolution by computing the eigenvector of the Liouvillian corresponding to the zero-eigenvalue. The results in Figure~\ref{fig:steady_xz_errors_vs_gammac} (lower left) show three distinct phenomena: (i) For low correction rates $\gamma_c$, energy is not removed fast enough to compete with heating, hence no significant cooling is achieved. (ii) For too high correction rates, gauge errors are removed too quickly and therefore do not have time to interact with the physical gauge sector by means of the unitary errors. Therefore cooling is also not observed in this ''quantum Zeno'' regime. (iii) In between these two extremes lies a regime where optimal cooling is achieved. If a gauge error occurs, it has time to interact with the physical sector and a subsequent correction removes its high energy contributions.
The steady-state gauge violation $\varepsilon_{ss}$ that is achieved scales mostly as $\varepsilon_{ss} \sim \frac{\gamma}{\gamma_c}$, as evident from Fig. \ref{fig:steady_xz_errors_vs_gammac} (upper left). Exceptions to this rule are regimes of weak correction where the gauge violation saturates as well as intermediate regimes with low incoherent error rates where coherent errors dominate the gauge violation.
Remarkably, these results are quite insensitive to the exact nature of the incoherent errors. In Figure~\ref{fig:steady_xz_errors_vs_gammac} (right) we show the same results for spontaneous emission $\sigma^-$ acting on all matter and link sites and the results still hold.
The dependence of these results on the magnitude of coherent errors and the gauge penalty is shown in the appendix.

\subsection{Stabilizing quantum simulations} 

\begin{figure}[h!]
    \centering
    \includegraphics[width=0.5\textwidth]{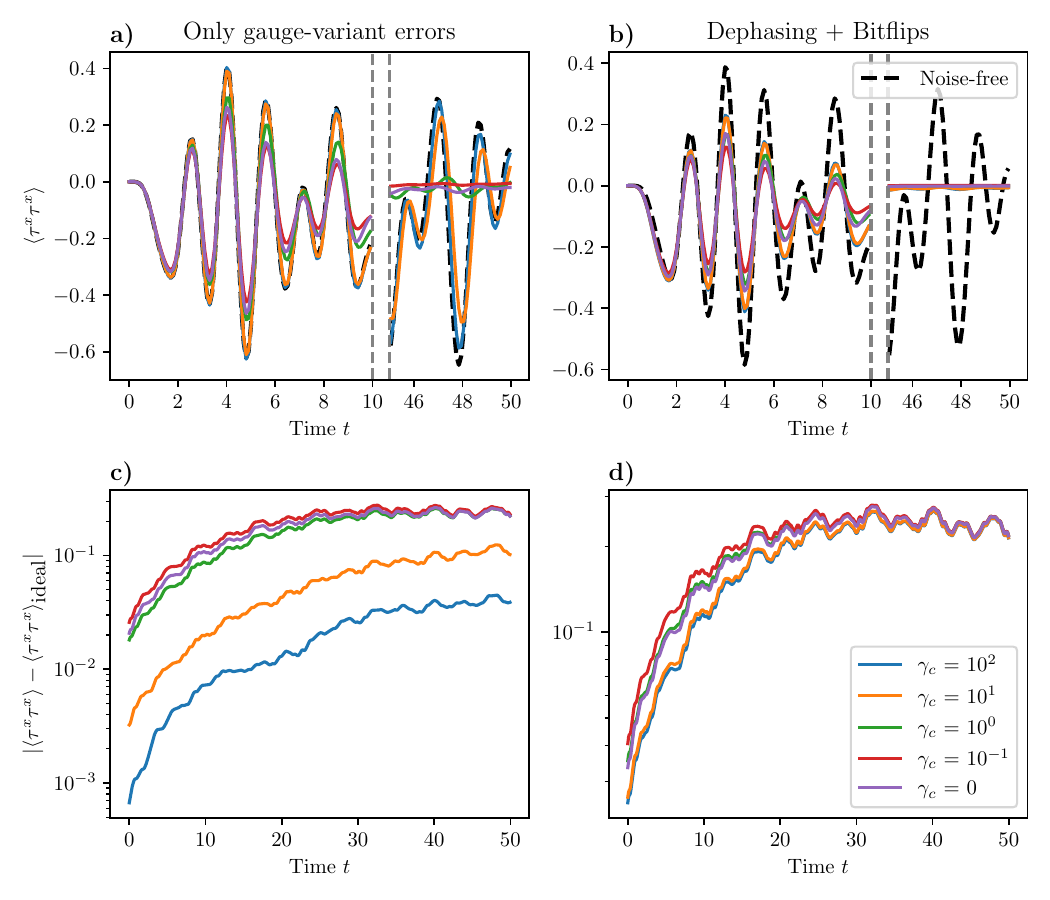}
    \caption{\textbf{a)} and \textbf{b)}: The link-link correlator \eqref{eq:link_correlator} is shown as a function of time for various correction rates $\gamma_c$. Without correction, the correlator decays, but the noise free dynamics can be restored by the correction scheme. \textbf{c)} and \textbf{d)}: Deviation from the noise free dynamics, time-averaged across 10 time units for improved legibility. ($\lambda=0.04, N=4$ and $\gamma=0.01$ for $\gamma_c = 0$ and $\gamma=0.01\cdot 11/8$ for $\gamma_c > 0$).}
    \label{fig:preserve_observable_dynamics}
\end{figure}

Next, we apply our scheme to increase the accuracy of observables
during a simulated time evolution. To this extent, we start the
system in a $J_a = 0$ eigenstate and then quench to the previously
used $J_a > 0$ and track the dynamics of observables over
time. Figure~\ref{fig:preserve_observable_dynamics} show the
trajectory of the observable
\begin{align}
    O = \sum_j \tau^x_{j-1, j}\tau^x_{j, j+1}, \label{eq:link_correlator}
\end{align}
i.e. a link-link correlator. In the appendix, we show results for more observables. The lower panels show the deviation of the observable from the exact, noise-free dynamics. This deviation is time-averaged over a running window of 10 time units for improved legibility. 

Previously we have treated the error rate $\gamma$ and the correction rate $\gamma_c$ as independent parameters. In practice, this will not be the case, since the gauge measurement and correction layers will themselves increase the error rate. We model this effect by equivalently writing noisy gates as a product of ideal gates and error terms in between. This allows us assume that the error rate is proportional to the number of CNOT operations executed per trotter step. This is a reasonable assumption, since in most experimental platforms, 2-qubit operations are the dominant source of errors \cite{hahn_integrated_2019, ryan-anderson_realization_2021}.
A single first-order trotter step requires 8 CNOT gates per site (as shown in the Appendix \ref{sec:cnot_count}), and the measurement of a single gauge operator requires 3 CNOT gates as seen in Fig.~\ref{fig:measurement_implementation}. In the worst case we need to do a layer of gauge measurements after every trotter step. This implies we perform $11/8$ more CNOT gates than in the case without gauge correction. When $\gamma_c \neq 0$, we assume this worst case and enhance the error rate $\gamma$ by this factor. In Appendix \ref{sec:trotter} we show using an example that this matches the noisy trotterized implementation.

We first focus on gauge-variant errors, i.e., bit-flips on matter sites and phase-flips on link sites. Since bit and phase errors can be exchanged by a local unitary transformation, this scenario is equivalent to the case where only bit or phase errors occur on all sites. Experimental platforms where one type of error dominates over the other are quite common, ranging from trapped ions \cite{ruster_long_lived_2016, vallabhapurapu_high-fidelity_2023} to solid-state spin qubits \cite{weimer_collectively_2013}.

As incoherent gauge-variant errors are localized single-qubit errors,
their presence causes in correlations to quickly decay to zero if no
correction is present, while the noise free evolution shows
interesting dynamics even for long times. As evident from
Fig.~\ref{fig:preserve_observable_dynamics} c), turning on
the gauge corrections increases the observable error if the correction rate is too small. But above a critical value of the correction rate, the scheme restores the original dynamics and prevents a decay of
the correlators. 
In particular at e.g. time $t=50$, the uncorrected observables are completely dephased, while the corrected counterpart shows an expectation value close to the noise-free case. Hence this shows a practical benefit of the correction scheme as extending the time-scale up to which local observables are reproduced correctly.
In this scenario with only gauge-invariant errors,
the gauge corrections represent a full error correction of individual
errors, as every error is visible as a gauge violation and can
uniquely be decoded and corrected. We note that this works most
reliably if the correction rate is faster than the timescale of the
Hamiltonian to be simulated. However, this is guaranteed in digital
quantum simulation approaches where the individual parts of the
Hamiltonian are implemented in a Trotterized form and the correction
is carried out after each Trotter step.

If we also turn on the gauge-invariant errors, the protective effect of the correction is reduced, but does not vanish entirely. The corrected dynamics still show reduced errors as compared to the uncorrected ones for sufficiently large correction rates. Since only the gauge-variant errors can be corrected, one may assume the simulation to be similar to an uncorrected simulation with only gauge-invariant errors. The effect of the correction is therefore to reduce the magnitude of the incoherent errors and hence increases the time-scale it takes observables to dephase. We therefore expect these results to hold regardless of the system size.

While sympathetic cooling requires moderate coherent errors $\lambda$ and comparatively small correction rates $\gamma_c$, stabilizing time evolution does not require coherent errors but higher correction rates.
Hence $\lambda$ and $\gamma_c$ are the crucial parameters that govern which of the two regimes are explored. 
Depending of the desired use-case, the coherent errors are either a result of simulation errors, or can be artificially added to engineer the cooling effect. Hence it is possible to tune both parameters to achieve the desired application.

\subsection{Discussion}

We have demonstrated that an active error correction scheme can successfully suppress gauge errors in lattice gauge theory simulations, inversely proportional to the correction rate. We showed how this gauge correction scheme can be interpreted as a sympathetic cooling setup and showed that this allows for tuning the parameters to the point where even ground states can be prepared and stabilized. This cooling scheme is very robust and its effects can be seen no matter the exact structure of the coherent errors, incoherent errors or the nature of the correction scheme. Our results can be directly applied to efforts to simulate lattice gauge theory on present NISQ devices \cite{Angelides2023}.
Finally we showed that this scheme can increase the accuracy of observables during simulation and therefore acts like a precursor to error correction.  It is particularly suitable in situations where errors are anisotropic as the scheme turns into full error correction in this case. However even without assumptions about the structure of the error, the scheme allows steady states of observables to be estimated more accurately.

\bibliography{references}

\textit{Note added: After submission of our manuscript, we
became aware of a related work employing similar techniques in order to post-select gauge-violating trajectories \cite{wauters_symmetry-protection_2024}.}
\begin{acknowledgements}
  Master-equation time-evolution trajectories and steady-state computations were computed using the \textsc{Qutip} library \cite{Johansson2013}. This work was funded by the Quantum Valley Lower Saxony (QVLS) through the Volkswagen foundation and the ministry for science and culture of Lower Saxony and by Germany’s Excellence Strategy – EXC-2123 QuantumFrontiers – 390837967.

  \end{acknowledgements}

\section*{Competing interests}
The authors declare no competing interests.
\appendix

\section{Parameter dependencies of steady state results}

\begin{figure}[h!]
    \centering
    \includegraphics[width=0.5\textwidth]{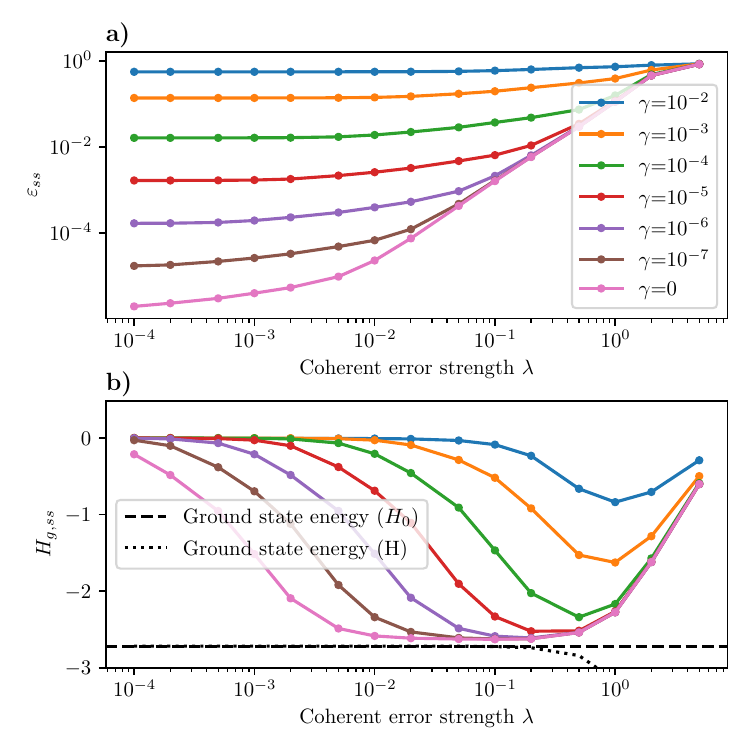}
    \caption{\textbf{a)} The steady state gauge violation \eqref{eq:gauge_violation} and \textbf{b)} gauge sector energy \eqref{eq:gauge_sector_energy} is shown for varying coherent error magnitudes $\lambda$ and spontaneous emission with rate $\gamma$ on all sites. ($\gamma_c=0.01, N=3$).}
    \label{fig:steady_state_vs_coherr}
\end{figure}

\begin{figure}[h!]
    \centering
    \includegraphics[width=0.5\textwidth]{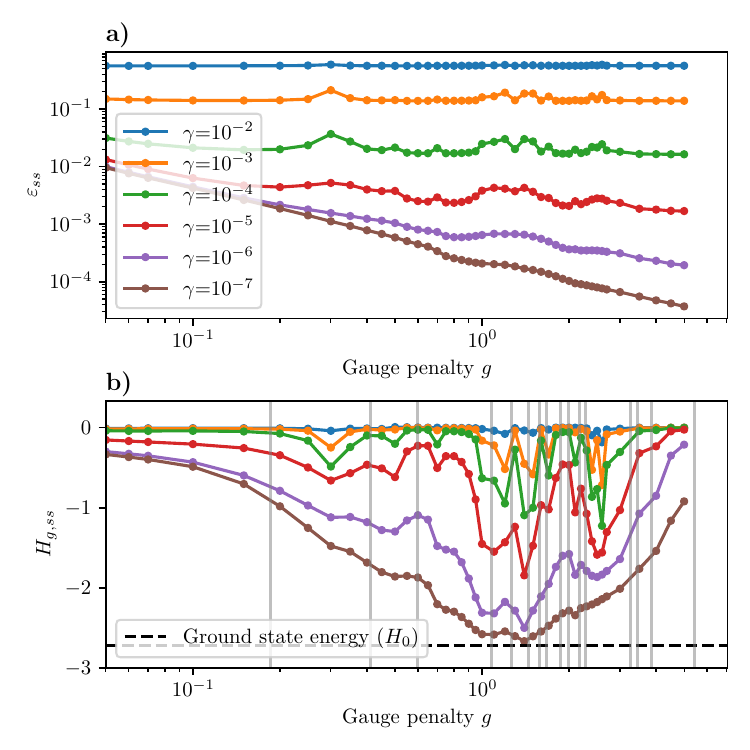}
    \caption{\textbf{a)} The steady state gauge violation \eqref{eq:gauge_violation} and \textbf{b)} gauge sector energy \eqref{eq:gauge_sector_energy} is shown for gauge penalties $g$ and spontaneous emission with rate $\gamma$ on all sites. Vertical gray lines indicate energy transitions in the system Hamiltonian. ($\gamma_c=0.01, N=3$).}
    \label{fig:steady_state_vs_penalty}
\end{figure}

Fig.~\ref{fig:steady_state_vs_coherr} shows the dependence of the achieved cooling on the magnitude of the coherent errors. Stronger decoherence requires more coherent errors, at the cost of a larger steady-state gauge violation. For smaller decoherence rates, the system is not sensitive to the exact value of the coherent error magnitude. 
In Figure~\ref{fig:steady_state_vs_penalty} we also show the dependence of the cooling results on the gauge penalty $g$. 
The best cooling is achieved when the gauge penalty lies in a region with many energy transitions of $H_0$ \cite{raghunandanInitializationQuantumSimulators2020}.
The steady-state gauge violations here recover the known result that the gauge penalty alone can yield a gauge-correcting effect \cite{halimeh_gauge_symmetry_2021} if only unitary errors are present.

\section{Matter-matter correlations}

\begin{figure}[!ht]
    \centering
    \includegraphics[width=0.5\textwidth]{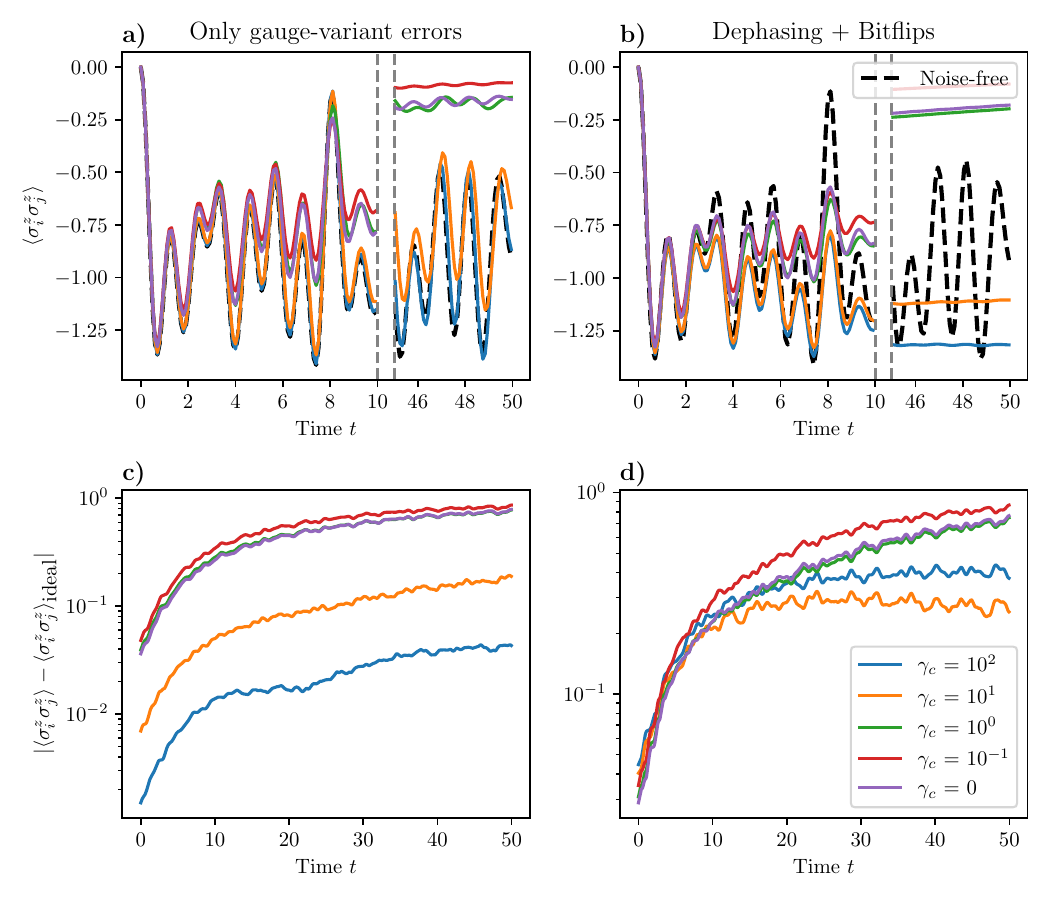}
    \caption{\textbf{a)} and \textbf{b)}: Matter-matter correlator is shown as a function of time for various correction rates $\gamma_c$. Without correction, the correlator decays, but the noise free dynamics can be restored by the correction scheme. \textbf{c)} and \textbf{d)}: Deviation from the noise free dynamics, time-averaged across 10 time units for improved legibility ($\lambda=0.04, N=4$ and $\gamma=0.01$ for $\gamma_c = 0$ and $\gamma=0.01\cdot 11/8$ for $\gamma_c > 0$).}
    \label{fig:preserve_observable_dynamics_2}
\end{figure}

In special cases, the steady-error of observables may even be reduced by the correction scheme as shown in Fig.~\ref{fig:preserve_observable_dynamics_2},  where a matter-matter correlator is tracked over time. As with the previous results, gauge-variant errors can be fully corrected. In contrast to the result in the main text, the long-term behavior of the uncorrected results starts to significantly deviate from the true solution, while the errors of the corrected simulations remain bounded.

\section{CNOT count for trotter implementation}\label{sec:cnot_count}

A trotterized implementation of the time-evolution of the system hamiltonian in Eq.~\eqref{eq:hamiltonian_base} requires a gate-level implementation of the three-qubit term 
\begin{align}
J_a(\sigma^+_j\taulink{j}^z\sigma^-_{j+1} + \text{h.c.}). \label{eq:local_hamiltonian_term}
\end{align}
This can be decomposed into Pauli-matrices as $J_a(\sigma^x\sigma^z\sigma^x + \sigma^y\sigma^z\sigma^y)$, acting on the sites $j$, $(j, j+1)$ and $j+1$. Time-evolution by time $\delta t$ of this term is given by 
\begin{align*}
\exp\left(\text{i}\delta t J_a (\sigma^x\sigma^z\sigma^x + \sigma^y\sigma^z\sigma^y)\right) \\
=\exp\left(\text{i}\delta t J_a \sigma^x\sigma^z\sigma^x\right)  \exp\left(\text{i}\delta t J_a\sigma^y\sigma^z\sigma^y)\right).
\end{align*}
Each of these two factors is equivalent up to local unitaries to $\exp\left(\text{i}\delta t J_a\sigma^z\sigma^z\sigma^z)\right)$, which can be implemented using the circuit \cite{Weimer2011} \\
\begin{center}
\begin{quantikz}
&\ctrl{2}&\qw     & \qw     & \qw     & \ctrl{2} & \qw\\
&\qw     &\ctrl{1}& \qw     & \ctrl{1}& \qw      & \qw\\
&\targ{} &\targ{} &\gate{\text{Rz}(2\delta t J_a)}& \targ{} & \targ{}  & \qw
\end{quantikz}
\end{center}
containing four CNOT operations and a single qubit rotation around the $z$-axis. Since we need two such operations for each of the local hamiltonian terms \eqref{eq:local_hamiltonian_term}, we end up with eight CNOT gates per trotter step and per gauge site.

\section{Trotterized implementation}\label{sec:trotter}

\begin{figure}[!ht]
    \centering
    \includegraphics[width=0.5\textwidth]{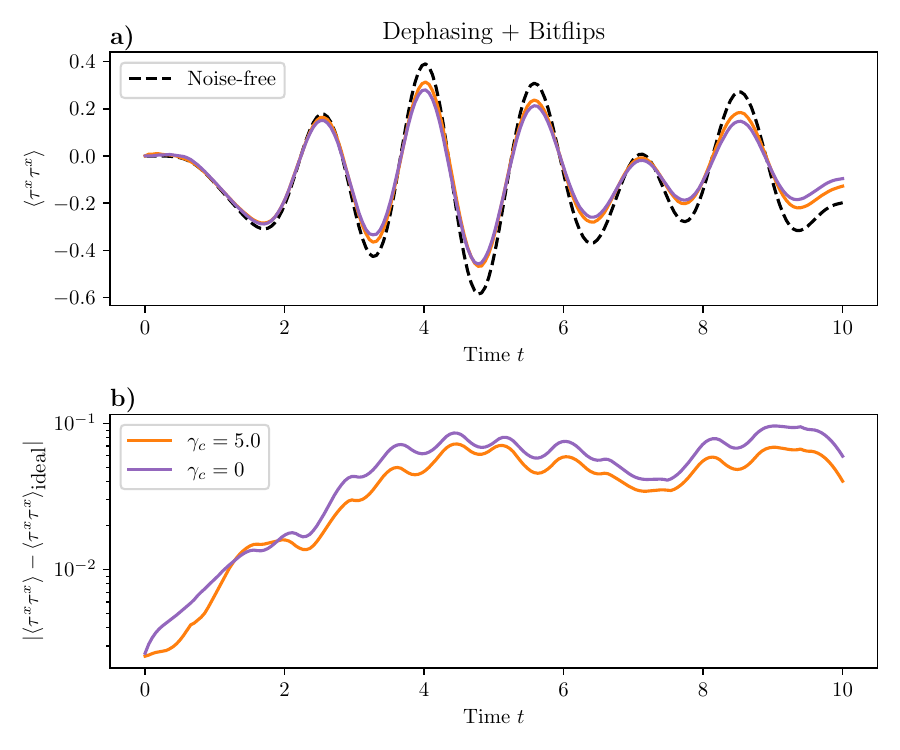}
    \caption{The results from Fig.~\ref{fig:preserve_observable_dynamics} are reproduced using a noisy Trotterized implementation. As with the continous-time implementation, an increase in observable accuracy is observed using the gauge-correction scheme. ($\lambda=0, N=4, \gamma=0.01$)}
    \label{fig:preserve_observable_dynamics_trotter}
\end{figure}

To verify that the continuous time-evolution results shown above can equivalently be obtained using an experimentally implementable Trotterized circuit, we here reproduce the results from Fig.~\ref{fig:preserve_observable_dynamics} b) and d). This is done by simulating the first-order Trotterized dynamics using a noisy, gate-based circuit. This circuit is converted to a basis consisting of CNOT and single-qubit operations. The CNOT gates are replaced by a quantum channel that has a probability $p$ of applying a bit- or phase-flip before or after the ideal gate, and a probability $1-p$ of executing an ideal CNOT. We fix the Trotter time-step $dt=0.05$ and select $p$ such that the resulting effective error rate per physical site 
\begin{align}
    \gamma &= 0.01 \frac{\text{Errors}}{\text{Simulated time}} \\ &= \frac{\text{\# Errors}}{\text{CNOT}}\cdot\frac{\text{\# CNOTs}}{\text{Trotter step}}\cdot \frac{\text{\# Trotter steps}}{\text{Simulated time}} \\
    &= p \cdot 8 \cdot \frac{1}{dt}
\end{align}
matches the order of magnitude used above. To match the correction rate per physical site 
\begin{align}
\gamma_c &= 5 \frac{\text{Corrections}}{\text{Simulated time}} \\
&= \frac{\text{\# Corrections}}{\text{Trotter step}} \cdot \frac{\text{\# Trotter steps}}{\text{Simulated time}},
\end{align}
we add a correction layer every four Trotter steps. As shown in Fig.~\ref{fig:preserve_observable_dynamics_trotter}, this scheme reproduces the continuous-time implementation, even in this case of gauge-variant and gauge-invariant errors and an improvement over the uncorrected case is again visible. 

\end{document}